# Spectroscopic thickness and quality metrics for PtSe$_2$ layers produced by top-down and bottom-up techniques

Beata M. Szydłowska[1,2], Oliver Hartwig[2], Bartlomiej Tywoniuk[2], Tomáš Hartman[4], Tanja Stimpel-Lindner[2], Zdeněk Sofer[4], Niall McEvoy[3], Georg S. Duesberg[2] and Claudia Backes[1*]

[1] Institute of Physical Chemistry, Heidelberg University, The Ruprecht Karl University of Heidelberg, D-69120 Heidelberg, Germany

[2] Institute of Physics, EIT 2, Faculty of Electrical Engineering and Information Technology Universität der Bundeswehr, 85579 Neubiberg, Germany

[3] AMBER & School of Chemistry, Trinity College Dublin, Dublin 2, Ireland

[4] Dept. of Inorganic Chemistry, University of Chemistry and Technology Prague, Technicka 5, 166 28 Prague 6, Czech Republic

*backes@uni-heidelberg.de

*Abstract*

Thin films of noble-metal-based transition metal dichalcogenides, such as PtSe$_2$, have attracted increasing attention due to their interesting layer-number dependent properties and application potential. While it is difficult to cleave bulk crystals down to mono- and few-layers, a range of growth techniques have been established producing material of varying quality and layer number. However, to date, no reliable high-throughput characterization to assess layer number exists. Here, we use top-down liquid phase exfoliation (LPE) coupled with centrifugation to produce widely basal plane defect-free PtSe$_2$ nanosheets of varying sizes and thicknesses. Quantification of the lateral dimensions by statistical atomic force microscopy allows us to quantitatively link information contained in optical spectra to the dimensions. For LPE nanosheets we establish metrics for lateral size and layer number based on extinction spectroscopy. Further, we compare the Raman spectroscopic response of LPE nanosheets with micromechanically exfoliated PtSe$_2$, as well as thin films produced by a range of bottom up techniques. We demonstrate that the $E_g^1$ peak position and the intensity ratio of the $E_g^1$/ $A_{1g}^1$ peaks can serve as robust metric for layer number across all sample types and will be of importance in future benchmarking of PtSe$_2$ films.

*Introduction*

Recently, the group-X transition metal dichalcogenides (TMDCs), referred to as noble-metal-based transition metal dichalcogenides (NTMDs), have been studied in their 2-dimensional form[1-4] and are gaining unprecedented interest with rapidly growing developments.[5-7] Their properties differ from the more widely studied group-VI TMDCs such as $MoS_2$ due to the fact that all d-orbitals are filled in NTMDs so that they tend to form $d^2sp^3$ hybridisation which stabilises the trigonal prismatic 1T crystal structure thermodynamically over the octahedral 1H-phase.[8] The non-bonded, filled d-orbitals and strong electronic hybridisation between adjacent layers makes it challenging to isolate few- and monolayer nanosheets of $PtSe_2$ micromechanically from bulk crystals.[9] However, exfoliation is particularly interesting, as it allows the electronic structure to be modulated by varying the layer number resulting in a tuneable bandgap.[1, 6, 9] $PtSe_2$ undergoes a semimetal to semiconductor transition when exfoliated down to < 3 layers as recently confirmed on high-quality samples.[10-12] Additionally, the NMTDs' band structure can be adjusted by external factors such as electric fields, doping[13] and stress or strain.[14] For example, under 8% strain, monolayer $PtSe_2$ undergoes a transition from an indirect-bandgap semiconductor to a direct-bandgap semiconductor.[9, 14, 15]

Moreover, $PtSe_2$ exhibits good air stability over elongated time periods of months[16] in contrast to for example black phosphorus.[17, 18] $PtSe_2$ is intriguing for electronic applications displaying carrier mobilities of ~200 $cm^2$ $V^{-1}$ $s^{-1}$ (based on FETs)[16] and strongly temperature-dependent on/off ratios of $10^3$-$10^8$ at 300 K and 50 K respectively, significantly larger than other TMDCs.[13, 19] Even though these numbers are already impressive, it is predicted that the carrier mobility of $PtSe_2$ can reach 1000 $cm^2$ $V^{-1}$ $s^{-1}$ at ambient temperature.[20] Overall, $PtSe_2$ has demonstrated application potential in various areas and was for example investigated in field effect transistors,[16, 21] wide spectral response photodiodes,[22] photodetectors,[23] catalysis,[11] and sensors.[24, 25]

Despite such progress, the understanding of the synthesis and properties of thin films of NTMDs is still in its infancy. Therefore, many challenges need to be addressed. For example, in order to study the layer-dependent properties in depth and establish structure-property relationships, accurate control of the material dimensions with simultaneous preservation of the high quality of the structure is required. In light of the application potential, large-scale production methods should be explored in order to meet future needs of the industrialisation,[7] where large quantities of uniformly-crystalline NTMDs will be a prerequisite. In addition to micromechanical exfoliation,[12] thin films of $PtSe_2$ have been produced by a range of bottom-up growth techniques including thermally-assisted conversion (TAC),[6, 10, 22, 24, 26-28] chemical vapour deposition (CVD),[2, 6, 11, 13, 29] chemical vapour transport (CVT)[6] and molecular beam epitaxy (MBE).[16, 30]

For further optimisation of the sample production, high-throughput characterisation techniques to assess both quality and layer number, for example based on optical spectroscopy, are extremely powerful.

It has already been shown that systematic changes in Raman spectra of PtSe$_2$ occur across samples yielded with different methods.[11, 26, 30] In particular, the E$_g^1$ mode shifts to higher wavenumber with decreasing thickness and changes in the relative intensity of the A$_{1g}^1$ and E$_g^1$ mode were observed. However, it is not clear if these trends originate from variation in thickness only or are affected by grain sizes and defectiveness. Hence, it would be helpful to clarify this unambiguously and to identify spectroscopic metrics to assess size and quality of the material. In order to do so, a detailed comparison of the PtSe$_2$ material produced by various techniques is required. In addition, it would be beneficial to produce nanosheets with well-defined thickness, but with as little basal plane defects as possible. Micromechanical exfoliation would be the top-down production technique of choice to achieve this. However, as mentioned above, in the case of PtSe$_2$, it suffers from the drawback that it is very difficult to obtain mono- and few-layered sheets.

Here, we suggest that top-down production by liquid phase exfoliation (LPE) can offer an alternative to bridge this gap. LPE is known as a versatile exfoliation strategy that can be applied to a broad range of layered bulk crystals[31-36] and to produce nanosheets with a relatively low content of basal-plane defects in dispersions. For group VI-TMDCs, this is manifested in a narrow exciton linewidth (~30-40 meV for monolayer WS$_2$ measured in room temperature luminescence).[37, 38] While colloidal dispersions originating from LPE contain nanosheets with characteristic broad lateral-size and thickness distributions, these can be narrowed by centrifugation-based size selection,[39] for example liquid-cascade centrifugation.[37] Such an approach can serve as advantageous to study size-dependent properties of the material by providing access to tailored lateral dimensions and thickness, without changing the structural integrity of the basal plane. Since an ensemble is always studied spectroscopically in LPE dispersions, inherent flake-to-flake variations are averaged out and the statistical mean is obtained. With knowledge of the lateral dimensions and thicknesses that can be obtained from atomic force microscopy (AFM) statistics, it is thus possible to establish quantitative metrics for lateral size and thickness as demonstrated for graphene,[40, 41] h-BN,[42] group VI-TMDCs,[43, 44] Ni$_2$P$_2$S$_6$,[45] black phosphorus,[46] GaS,[47] layered hydroxides[48] to name a few.

In this work, we perform LPE of PtSe$_2$ in aqueous surfactant and isolate fractions of varying size and thickness distributions that we quantify by statistical AFM. We identify size-dependent changes in optical absorbance and extinction spectroscopy that we quantitatively link to nanosheet lateral size and layer number, respectively. In addition, we carry out a detailed analysis of the Raman spectroscopic fingerprint and compare the LPE samples to a range of PtSe$_2$ nanosheets and films produced by mechanical cleavage and bottom-up growth, where the Raman spectra were digitized from literature and analysed in the same manner. With this, we identify universal Raman spectroscopic metrics for layer number and sample crystallinity/defectiveness that can be applied to PtSe$_2$ nanosheets produced by numerous techniques.

## Results and discussion

### Liquid Phase Exfoliation

For LPE, PtSe$_2$ bulk crystals were synthesised from the elements (see Methods) and subjected to probe sonication. The energy input overcomes the attraction between the layers resulting in exfoliation (and scission).[49] Colloidal, stable dispersions from LPE can be typically obtained when nanosheet aggregation is prevented by appropriately chosen solvents with matching solubility parameters[34, 50, 51] or surfactants in aqueous solution.[34] Due to the reported good environmental stability of PtSe$_2$,[16] we chose the surfactant sodium cholate as stabiliser in aqueous medium, as it was reported that a larger population of thin nanosheets can be obtained compared to solvent-based exfoliation.[49] LPE is a relatively crude method that yields polydisperse samples of nanosheets with broad lateral-size and thickness distributions.[49] This can also be regarded as advantage, as it is possible to isolate size-selected fractions from the same stock, for example by liquid cascade centrifugation (LCC).[37] This procedure involves centrifugation of the as-obtained stock dispersion with iteratively increasing centrifugal accelerations. After each step, sediment and supernatant are separated, the sediment is collected for analysis, while the supernatant is subjected to centrifugation at higher speeds. Here, we performed a 6-step LCC with increasing centrifugal accelerations (expressed as relative centrifugal force, *RCF*, in units of the earth's gravitational field) between 100 $g$ and 30,000 $g$ to size-select the dispersion resulting in five PtSe$_2$ fractions with relatively large/thick nanosheets isolated at low accelerations and small/thin nanosheets at high accelerations. Fractions were labelled by the lower and upper centrifugation boundary used for preparing the respective fraction (for details see Methods).

The nanosheets contained in the dispersion were drop-cast on Si/SiO$_2$ wafers and AFM was used to measure the lateral dimensions and thickness in isolated fractions. Representative images for the fraction of large/thick nanosheets isolated at 0.1-0.4 k $g$ and small/thin nanosheets at 10-30 k $g$ are displayed in Figure 1A-B. For all data, see Supporting Information (SI), Figure S1. For each fraction about 200 individual nanosheets were imaged and their length (L, longest dimension), width (W, perpendicular to L) as well as thickness (T) extracted. A semi-automated analysis was used similar to the report by Fernandes *et al.* Fernandes T.F.D.;Miquita D.R.;Soares E.M.;Santos A.P.;Cançado L.G.Neves B.R.A. [52] and our previously reported calibration to account for cantilever broadening and pixellation effects in the determination of lateral dimensions.[53] Furthermore, it was considered that the apparent AFM height of monolayer nanosheets is typically larger when compared with the theoretical thickness of a given material.[43, 49, 54, 55] This is due to intercalated and/or adsorbed water/surfactant. In analogy to previous reports,[42, 43, 46, 49, 54, 56] we used step-height analysis to determine the apparent thickness of the LPE PtSe$_2$ monolayer seen by AFM. Multiple steps and terraces were measured (see SI, Figure S2) that were found to be a multiple of 2.05 nm which is thus considered as the apparent AFM thickness of one layer of LPE PtSe$_2$ and used to convert the thickness to layer number, N.

With this information, size distribution histograms can be constructed, for example L and N histograms for fractions 0.1-0.4 k $g$ and 10-30 k $g$ shown in Figure 1C-D. For all fractions the histogram shape follows a lognormal distribution with mean sizes decreasing as the central accelerations increase as expected. For additional histograms see SI, Figure S3-4. From this data, the average lateral dimensions (<L>) and layer number (<N>) can be calculated. The mean nanosheet length is plotted versus central *RCF* in Figure 1E, while the mean layer number is shown in Figure 1F. The sizes decrease as a power law with increasing centrifugal force, which is characteristic for the centrifugation process.[57] It is clear that the sheet distributions become narrower, and sheets are simultaneously smaller and thinner for increasing centrifugal accelerations. The mean length ranges from ~ 30- 170 nm, while the layer number spans from ~2-10.

The data in Figure 1E-F suggest that nanosheet lateral size and thickness cannot be easily varied independently. While it can be anticipated that the link between <L> and <N> is due to the centrifugation that predominantly separates material by mass/density, it was recently shown that this relationship is governed by the fundamentals of the exfoliation process, where both exfoliation and scission occur with an equipartition of energy.[49] To illustrate this, Figure 1G shows a scatter plot of the nanosheet area (estimated as L * W) as a function of layer number. Each data point in Figure 1G represents an individual nanosheet in the stock dispersion produced via LPE. Each fraction is depicted as a different colour with blue and black corresponding to fraction 0.1-0.4 k $g$ and 10-30 k $g$, respectively. The plot demonstrates that the correlation of <L> and <N> in the fractions after size selection is not (only) due to centrifugation, but a consequence of the initial population and that nanosheets with larger area tend to be thicker while smaller nanosheets tend to be thinner in agreement with other layered materials.[49] Overall, the AFM analysis confirms that a relatively wide spread of lateral sizes and thicknesses are accessible. Note that an average layer number of ~2 in the 10-30 k $g$ fraction is thinner than what is commonly achievable by micromechanical exfoliation so that the LPE samples should be well suited for studying size-dependent optical properties.

*Spectroscopic Metrics for Size and Thickness of LPE Nanosheets Based on Extinction and Absorbance Spectra*

Prior to establishing spectroscopic metrics, it is important to verify the structural integrity of the nanosheets. To this end, X-ray photoelectron spectroscopy (XPS) was performed on the fractions of largest (0.1-0.4 k $g$) and smallest (10-30 k $g$) nanosheets, as well as the initial bulk material. The fitted XPS core-level spectra of the smallest sheets are shown in Figure 2A-B, while spectra corresponding to large nanosheets and bulk material are provided in the SI, Figure S5. In all cases, platinum core levels (Pt 4f) can be fitted with 3 doublets pointing to the presence of various Pt (II) and Pt (IV) species. The dominant signal (blue; binding energy 73.4 eV and 76.7 eV) is assigned to $PtSe_2$. In addition to this, doublets at both higher and lower binding energies are observed. The signal at lower binding energy (light blue; 72.2 eV and 75.6 eV) is attributed to Pt(II) species for example due to selenide-deficient structures ($PtSe_x$).

Additionally another Pt(IV) doublet at higher binding energy (orange; 74.3 eV and 77.7 eV) reflects a small portion of oxidised material (likely $PtO_2$). All signals are also present in the large nanosheets and the bulk material, albeit with lower relative intensities of the Pt(II) and oxide species. Based on the presented spectra and their fitting, the composition of the $PtSe_2$ samples was evaluated. The samples show an oxide content of 7, 11 and 14% for bulk, large and small $PtSe_2$ respectively confirming the high stability of the material despite exposure to oxygen and humidity. This conclusion is supported by fitting of the selenium (Se 3d) core levels (see Figure 2B and SI, Figure S5). The selenium core levels can be well described by two doublets. The dominant peaks at 54.7 and 55.6 eV are attributed to selenium in $PtSe_2$, while the doublet at higher binding energy is assigned to sub-coordinated selenium. Since the relative intensity of this signal increases with decreasing size, it is likely related to nanosheet edges in agreement with previous reports, where it was assigned to grain boundaries in TAC-grown $PtSe_2$.[24, 28] Overall, the XPS analysis confirms a low degree of defectiveness which is a prerequisite to study intrinsic size and thickness-dependent properties.

It is known from a range of 2D materials that the optical extinction and absorbance spectra change systematically with the nanosheet dimensions.[40-48, 56, 58] Hence, with knowledge of nanosheet size and thickness, quantitative spectroscopic metrics can be established. To do so for $PtSe_2$, we measure the extinction spectra of the size-selected dispersions in a standard transmission mode and the absorbance with the cuvette placed in the centre of an integrating sphere, where scattered light is collected. It should be noted that extinction spectra of nanomaterials are the sum of the absorption and scattering.[43, 48, 59] Thus, to access material-dependent changes of the optical properties, it is beneficial to perform both measurements. Normalised extinction and absorbance spectra of the size-selected $PtSe_2$ are displayed in Figure 2C-D. As expected, the extinction spectra (Figure 2C) show a clear size dependence. For each fraction, there are two main features visible: a relatively narrow peak at ~400 nm and a less defined shoulder at ~800 nm. For the smallest $PtSe_2$ nanosheets, the spectrum is dominated by the peak at 400 nm. As the nanosheet size increases, the shoulder at ~800 nm emerges and gets more pronounced. In the 0.1-0.4 k $g$ fraction, containing the largest nanosheets, the spectral profile is very broad and dominated by the shoulder at ~800 nm. Since the wavelength-dependent scattering spectra follow the absorbance in shape (with a red-shift),[43, 59] it can be anticipated that the spectral broadening arises from an increasing contribution of scattering as the nanosheets get larger. However, the absorbance measurements performed in the integrating sphere (Figure 2D) show that the absorbance spectra are virtually identical to the extinction. This suggests that the scattering contribution is minor in the $PtSe_2$ dispersions pointing to a high oscillator strength. This is confirmed by the direct comparison of extinction and absorbance shown in the SI, Figure S6 which illustrates that scattering is indeed negligible. At high wavelength (> 1100 nm), the absorbance of the smallest/thinnest nanosheets approaches zero, which is consistent with the reported semi-metal to semiconductor transition.[10-12]

In analogy to group-VI TMDCs,[37, 43, 44] we interpret the size-dependent changes in terms of edge and confinement effects. Edges are electronically different from the basal plane and hence possess a different absorbance/extinction coefficient which leads to a variation in the relative oscillator strength that can be expressed by analysing intensity ratios at different wavelengths. In addition, we expect confinement and dielectric screening effects of excitons to cause shifts in peak positions. To test whether similar effects occur in PtSe$_2$, we plot the intensity ratio of the local minimum and the intensity of the upcoming shoulder $Ext_{255}/Ext_{800}$ and $Abs_{225}/Abs_{800}$ as a function of mean nanosheet length <L> in Figure 2E. Due to the negligible scattering, we find an excellent agreement between absorbance and extinction spectra across the entire size range. According to the model, assuming a different extinction/absorbance coefficient at the edge and centre of the basal planes,[43] we expect the data to follow Equation 1, with x as the thickness of the edge region and k being the nanosheet aspect ratio (length/width).

$$\frac{Abs_{(\lambda_1)}}{Abs_{(\lambda_2)}} = \frac{\alpha^{center}_{(\lambda_1)}<L>+2x(k+1)\Delta\alpha_{(\lambda_1)}}{\alpha^{center}_{(\lambda_2)}<L>+2x(k+1)\Delta\alpha_{(\lambda_2)}} \qquad \text{Equation 1}$$

The line in Figure 2E is a fit to Equation 1 which describes the data very well. Rearranging the equation with the fit values obtained allows us to establish a quantitative metric to extract the average lateral dimensions of PtSe$_2$ nanosheets from extinction (or absorbance) spectra according to Equation 2.

$$<L>[nm] = \frac{0.01335\frac{Ext255}{Ext800}-0.06}{0.00354\frac{Ext255}{Ext800}+1} \qquad \text{Equation 2}$$

In addition, we extract the peak position of the main peak at 300-500 nm from the second derivative spectra (SI, Figure S7), where a clear peak shift is observed (even though the peak is broad). This is attributed to confinement and dielectric screening effects so that the result is plotted as function of average layer number, <N>, in Figure 2F. We find an exponential red-shift of the peak with increasing layer number in agreement with group-VI TMDCs.[44] Again, data from absorbance and extinction coincides. Fitting allows us to formulate Equation 3 that relates the peak position (E in eV) to average layer number:

$$<N> = -\ln\left(\frac{E-3.185}{0.3157}\right)*3.912+1 \qquad \text{Equation 3}$$

Such an understanding of the optical absorbance is thus of great use to extract lateral size and thickness of nanosheets in any unknown PtSe$_2$ dispersion. In addition, we expect that similar changes to the spectral profile can be observed in reflectivity measurements on substrate-supported sheets as was demonstrated previously for group-VI TMDCs.[44, 58] Last but not least, optical extinction and absorbance offer the possibility to investigate the environmental stability of the material, as degradation will reduce the extinction/absorbance intensity. The PtSe$_2$ nanosheets were confirmed to be environmentally stable after LPE and size-selection procedures, both carried in the aqueous environment, as the extinction spectra of the dispersions recorded over a period of weeks exhibited virtually no changes (see SI, Figure S8).

*Spectroscopic Thickness and Quality Metrics for Various Types of PtSe$_2$ Based on Raman Spectroscopy*

While metrics based on absorbance are undoubtedly useful, it would be of greater relevance to establish metrics based on a spectroscopy techniques that is more accessible for substrate-supported samples, as these could then also be readily applied to bottom-up synthesised material. Raman spectroscopy holds promise in this regard, in particular because systematic changes across PtSe$_2$ samples have already been observed in other reports.[11, 26, 30] To test whether such changes can also be observed in PtSe$_2$ fabricated by top-down techniques, Raman spectra of all five fractions of LPE samples were measured at low laser power (~1.23 μW) to avoid heating after dropcasting the samples on Si/SiO$_2$. Normalised LPE PtSe$_2$ spectra shown in Figure 3A are dominated by the two characteristic modes: $E_g^1$ and $A_{1g}^1$ at around ~179 cm$^{-1}$ and ~207 cm$^{-1}$ respectively. Similar to other materials, we observe size-dependent changes in Raman spectra of PtSe$_2$. These include an upshift of the $E_g^1$ mode by ~2 cm$^{-1}$ as the sheets get thinned down from bulk to ~2 layers and a simultaneous significant decrease of the relative intensity of the $A_{1g}^1$ mode with the layer number; from 0.33 to 0.63.

First, we compare these changes to PtSe$_2$ obtained from micromechanical exfoliation (referred to as "Tape", see SI, Figure S9-10). Not only are the trends well reproduced in the Raman spectra displayed in Figure 3B, but the changes are also similar quantitatively. To analyse the data further, the Raman spectra were fit with Lorentzian functions in the spectral region of interest (see SI Figure S11) to extract the $E_g^1$ and the $A_{1g}^1$ position, their respective intensity ratio (I $E_g^1$/I $A_{1g}^1$) and the full width at half maximum (FWHM) of the $E_g^1$ mode. Intuitively, based on the knowledge of group-VI TMDCs,[60, 61] one would expect the peak position to be sensitive to the layer number. To test this, we plot the position of the $E_g^1$ vibration as a function of inverse layer number (1/<N>) in Figure 3C. This representation allows us to include the bulk data point in the x-axis origin. A clear scaling of the peak position with layer number is observed (as indicated by the line, which will be discussed further below) with samples produced from LPE and micromechanical exfoliation showing only little deviation from each other.

To test whether this is a universal behaviour, we further measured and analysed the Raman spectra of TAC-grown samples (Figure 3D). Although, the trends are in agreement, the changes are of substantially different magnitude. For instance, for the LPE and Tape samples, the $E_g^1$ shifts by ~2 cm$^{-1}$ and the $A_{1g}^1$/$E_g^1$ intensity ratio changes between 0.7 and 0.3 overall. In case of the TAC sample, the $E_g^1$ shifts up to 7 cm$^{-1}$ and the $A_{1g}^1$/$E_g^1$ intensity ratio changes between 1.16 and 0.23, i.e. much more substantially. In addition, the vibrational modes in the TAC samples are visually significantly broadened which could be an indicator for a lower degree of crystallinity compared to the PtSe$_2$ from top-down techniques.

In order to evaluate the above mentioned changes further, all the recorded spectra as well as Raman spectra digitized from the literature (SI, Figure S12-S17)[2, 6, 11, 13, 16, 29, 30, 62] were fitted with 2 Lorentzian functions to extract positions, widths and intensity ratios. In Figure 3E-H all extracted parameters are plotted collectively, regardless of the material fabrication method. In Figure 3E, the $E_g^1$ position is plotted as a function of inverse layer number. Across all studied samples, it shifts between 172 cm$^{-1}$ and 183 cm$^{-1}$, i.e. over a much larger range than the shift observed for the LPE and micromechanically exfoliated samples. Note, the line from Figure 3C is included as a guide for the eye in Figure 3E. This suggests that sample quality (e.g. crystallinity, grain size, degree of oxidation, structural perfection in general) might contribute to the peak position in addition to thickness.

In an attempt to access the average sample quality by Raman spectroscopy, we analysed the FWHM of the $E_g^1$ mode. This is plotted as a function of inverse nanosheet thickness in Figure 3F. The data is overall very scattered, showing a spread from ~3-17 cm$^{-1}$ across all samples. No clear scaling with nanosheet thickness is observed except for the LPE samples (black stars), where the peak width increases with decreasing layer number (as would be expected for an increasing level of defectiveness when sheets become smaller/thinner). Despite this broad range (~15 cm$^{-1}$) of the $E_g^1$ FWHM, the majority of samples are characterised by a FWHM of 3-7 cm$^{-1}$ (shaded area in Figure 3F). With this observation, we tentatively define the $E_g^1$ FWHM of 7 cm$^{-1}$ as a quality indicator. Clearly, epitaxially-grown PtSe$_2$ from MBE[30] shows a very good quality with a narrow linewidth even down to the monolayer. Note that the linewidth can also be dependent on the grating used for the Raman measurements. However, no trend could be identified.

When we exclude all data with an $E_g^1$ FWHM > 7 cm$^{-1}$ and reconsider the evaluation of the $E_g^1$ position as function of inverse nanosheet layer number, the plot, shown in Figure 3G, becomes significantly less scattered and a scaling behaviour is observed independent of the production technique. This plot is interesting for a number of reasons. Firstly, in retrospect, it supports choosing an $E_g^1$ FWHM of 7 cm$^{-1}$ as a quality indicator. Secondly, it confirms that LPE nanosheets are widely defect-free (except edges), since the scaling of the $E_g^1$ is similar to the micromechanically exfoliated samples. Note that the line in Figure 3C is actually the fit line from the collective data set shown in Figure 3G. Thirdly, and most importantly, it allows us to derive a quantitative spectroscopic metric for the thickness of PtSe$_2$, regardless of the production technique, with most data points falling within 20% error with respect to N (shaded areas in Figure 3G), as long as the criterion of $E_g^1$ FWHM > 7 cm$^{-1}$ is fulfilled. We can empirically fit the data to a power law. Rearranging for layer number gives Equation 4 that quantitatively links layer number to the Raman $E_g^1$ peak position (Pos$_{Eg}$ in cm$^{-1}$):

$$<N> = \left(\frac{Pos_{Eg}-177.7}{2.535}\right)^{-2.58} \qquad \text{Equation 4}$$

Finally, we address the relative change of the intensity of the $A_{1g}^1$ mode that was also observed to vary systematically with layer number. In Figure 3H, the intensity ratio of $E_g^1$ and $A_{1g}^1$ is plotted versus the inverse layer number. This plot includes the data from various production techniques, irrespective of the width of the vibrational modes, and is significantly less scattered than the $E_g^1$ position data shown in Figure

3E. This suggests that the peak intensity ratio can be used as a metric for layer number irrespective of production method and sample quality. While this appears at first glance to be extremely powerful, on the one hand, we would like to point out that the changes are more significant for thicker nanosheets, gradually levelling off for few-layered material (at large 1/N) thus potentially introducing a larger uncertainty when determining the layer number of very thin specimens. On the other hand, in support of using intensity ratios as foundation for metrics, it should be noted that peak intensity ratios are easily accessible without data fitting thus increasing their practical relevance. The data can again be fitted empirically with a power law (solid line in Figure 3H). Most data falls within 20 % uncertainty (grey shaded area). Rearranging gives the relation of layer number to intensity ratio of the $E_g^1$ and $A_{1g}^1$ ($E_g / A_{1g}$) in Equation 5 which is widely independent of production technique and sample quality.

$$<N> = \left( \frac{\frac{E_g}{A_{1g}} - 1.557}{-1.56} \right)^{-4.48} \qquad \text{Equation 5}$$

## *Conclusion*

In summary, we have used liquid phase exfoliation to produce nanosheets of PtSe$_2$ down to an average layer number of 2. The material has a low basal plane defect content, with only minor oxidation visible in XPS and Raman $E_g^1$ linewidths < 7 cm$^{-1}$. Size selection by cascade centrifugation and statistical size/thickness assessment has enabled us to quantitatively link size-dependent optical changes in absorbance and extinction to the nanosheet dimensions resulting in metrics to extract both lateral dimensions and layer number of the LPE nanosheets from extinction spectra.

Moreover, thickness dependent changes were observed in Raman spectroscopy with spectra from micromechanical exfoliation and LPE agreeing very well quantitatively confirming the high structural integrity and low degree of defectiveness of the LPE samples. As such, LPE samples are suitable to establish quantitative spectroscopic metrics. In the case of PtSe$_2$, they bridge the gap between relatively thick, high-quality micromechanically exfoliated nanosheets and thin films grown by various techniques with varying quality and crystallinity. To this end, we also analysed Raman spectra of PtSe$_2$ produced from a range of bottom-up techniques including TAC, CVD and MBE. Some of the data was digitized from existing literature and fitted to extract values of $E_g^1$ and $A_{1g}^1$ position, $E_g^1 / A_{1g}^1$ intensity ratio and FWHM of $E_g^1$. This collective and comparative study showed that the FWHM of $E_g^1$ mode seems to be a good indicator for the structural integrity of the samples. For samples with a linewidth of 3-7 cm$^{-1}$ the $E_g^1$ position scales with the layer number irrespective of the production technique. However, for samples with a linewidth > 7 cm$^{-1}$ a deviation is observed strongly suggesting that these are samples of lower crystallinity which has an additional impact on the position of the Raman modes. In contrast, the $E_g^1 / A_{1g}^1$ intensity ratio shows a rather robust scaling with layer number irrespective of linewidth. For both Raman position and intensity ratio, quantitative metrics are presented.

Thus, Raman can be used as a powerful characterization technique for $PtSe_2$ giving quantitative information on layer number and in addition a qualitative indication of sample crystallinity. Due to the ease of the measurement, this will be of uttermost importance in future work enabling benchmarking of $PtSe_2$ materials produced across the world, regardless of the fabrication method. We believe this will help to establish a common standard in the community, facilitate optimisation of growth conditions and will be helpful to establish structure-property relationships for $PtSe_2$ (also in applications).

## *Materials and Methods*

*Synthesis of $PtSe_2$.* Platinum sponge (99.99%) was obtained from Surepure Chemetals (USA) and selenium (99.9999%, granules 2-4 mm) from Wuhan Xinrong New Materials Co. (China). For the synthesis, platinum sponge and selenium granules corresponding to 2 g of $PtSe_2$ were placed in a quartz glass ampoule (25x100 mm, wall thickness 3 mm). Selenium was used in 2 at. % excess. The ampoule was evacuated to a pressure of $1 \times 10^{-3}$ Pa and sealed by an oxygen-hydrogen welding torch. Subsequently, the ampoule was heated on 1270 °C using a heating rate of 5 °C/min. After 30 minutes at the final temperature, the ampoule was cooled to 1000 °C using a cooling rate of 1 °C/min and finally freely cooled to room temperature inside the furnace. The selenium excess condenses on the opposite site of the ampoule and the formed crystalline block of $PtSe_2$ was removed from the ampoule. This $PtSe_2$ was used for LPE.

*Micromechanical exfoliation of $PtSe_2$.* Few-layer single crystal $PtSe_2$ flakes were transferred onto a clean silicon chip using the micromechanical exfoliation approach.[63] A $PtSe_2$ single crystal was purchased from HQ Graphene and its small portions peeled onto sticky tape, similar to scotch tape. By covering the material with another strip of tape and subsequently separating the tapes the crystal material was peeled-off. Repeating this process several times yields significantly thinned down crystal flakes on the tape. The tape is then placed onto a clean silicon chip and previously exfoliated crystal material (in contact with the silicon surface) is transferred. This leaves a random distribution of $PtSe_2$ flakes on the silicon surface which are later analysed using an optical microscope. In order to evaluate layer number, apparent thickness of the film was measured with AFM and divided by 0.6 nm corresponding to thickness of a single layer of $PtSe_2$.[30]

*Liquid phase exfoliation and size selection of $PtSe_2$.* The $PtSe_2$ crystal (0.5 g $L^{-1}$) was immersed in 35 mL of aqueous SC solution ($C_{surf}$ = 1.7 g $L^{-1}$). The mixture was sonicated under cooling in a metal beaker by probe sonication using a solid horn probe tip (Sonics VX-750) for 7.5 h at 30% amplitude with a pulse of 6 s on and 4 s off. The dispersion was kept in a 4 °C cooling cell during sonication to avoid heating. Samples prepared according to this protocol are referred to as stock dispersion.

Dispersions were size-selected by liquid cascade centrifugation[37] in multiple steps of subsequently increasing centrifugation speeds at 100 $g$, 400 $g$, 1,000 $g$, 3,000 $g$, 5,000 $g$, 10,000 $g$ and 30,000 $g$. After each step, the sediment was redispersed for analysis in a reduced solvent volume (~1 mL of 0.1 g/L SC) and the supernatant was used for the subsequent step. The sediment after the first step at 100 $g$ containing unexfoliated material was discarded, as well as the supernatant after 30 k $g$ containing impurities and very small sheets. A Hettich Mikro 220R centrifuge was used equipped with two different rotors: for centrifugation <5 k *g* a fixed-angle rotor 1016 and 50 mL vials (filled with 20 mL each); for centrifugation >5 k *g*, a fixed-angle rotor 1195-A and 1.5 ml vials. All centrifugation steps were performed for 2 h at 13 °C. The data and annotations in all Figures use the central *g*-force to express the consecutive centrifugation speeds. For example; the central *g*-force for a 0.4-1 k *g* sample (sediment collected after centrifuging at 1 k *g* supernatant after 0.4 k *g* step) would be 0.75 k *g*.

*Growth by thermally assisted conversion.* Pt thin films of varying thicknesses were sputtered from a MaTeck Pt target using a Gatan precision etching and coating system onto silicon (Si) substrates with 300 nm dry thermal oxide ($SiO_2$). The deposition rate and film thickness were monitored with a quartz-crystal microbalance. These films were then placed in the centre of a quartz-tube furnace and heated to a growth temperature of 450 °C under 150 sccm of 10% $H_2$/Ar flow. Selenium (Se) vapor was then produced by heating Se powder (Sigma Aldrich, 99.99%) to ~220 °C in an independently controlled, upstream heating zone of the furnace, and carried downstream to the Pt films for a duration of 2 h, after which the furnace was cooled to room temperature. Typically, the $PtSe_2$ films are by a factor of ~3.5 thicker than the initially sputtered Pt layer.[21, 26] Thus, to estimate the layer number of the TAC films, the initial Pt thickness was multiplied by a factor of 3.5 and divided by 0.6 nm which is the thickness of one TAC $PtSe_2$ layer.

*Atomic Force Microscopy*. A Bruker Icon Dimension atomic force microscope in ScanAsyst mode with Bruker OLTESPA-R3 cantilevers was used for all AFM measurements. Each dispersion was diluted with deionised water to an optical density ~0.2 at the local maximum and drop casted onto a preheated (160 °C) Si/$SiO_2$ (300 nm oxide layer) wafer, and further cleaned with deionised water and isopropanol to remove excess surfactant. Individual nanosheets were analysed as described previously[49] and the tip broadening effects were addressed by previously established length correction.[53] The apparent AFM heights of materials obtained *via* LPE are usually overestimated. To convert the apparent measured AFM thickness to the actual number of layers, we have applied an approach termed step height analysis: we measured the height of steps associated with terraces of incompletely exfoliated nanosheets on the nanosheet surface. In total 99 height profiles of incompletely exfoliated nanosheets were extracted. Values corresponding to the apparent height difference are plotted in ascending order in the SI. The apparent height of each step is a multiple of a value representing the apparent height of one $PtSe_2$ monolayer seen in the AFM. The theoretical per monolayer lattice constant is always much smaller. The step height of one LPE $PtSe_2$ monolayer was determined to be 2.05 nm. Therefore, to convert the measured apparent AFM height to number of layers, the measured height was divided by 2.05 nm throughout this manuscript.

*UV-Vis Spectroscopy.* Optical extinction and absorbance measurements were carried out in an Agilent Cary 6000i spectrometer in quartz cuvettes of 4 mm pathlength with surfactant concentrations of 0.1 g/L. The samples were diluted to optical densities below 0.2 in extinction across the entire spectral region. For the absorbance measurements, the spectrometer was equipped with an integrating sphere (external DRA-1800) and the cuvette was placed in the centre of the sphere. The measurements of both extinction and absorbance spectra allow for the calculation of scattering spectra (Sca=Ext−Abs). In both cases, the spectra were measured with 0.5 nm increments.

*Raman Spectroscopy of LPE samples.* Raman spectroscopy was carried out on a Renishaw InVia-Reflex confocal Raman microscope with a 532 nm excitation laser in air under ambient conditions. The Raman emission was collected by a 50x long working distance objective lens in streamline mode and dispersed by a 2400 l/mm grating with 1% of the laser power (1.23 µW). The spectrometer was calibrated to a silicon reference sample prior to the measurement to correct for the instrument response. Liquid dispersions were drop-casted (~30 µL) onto preheated (120 °C) Si/SiO$_2$ wafers and cleaned with deionised water and isopropanol. For each sample 10 spectra on different positions were recorded and averaged. After averaging, the spectra were baseline corrected in the region of 150-230 cm$^{-1}$ and fitted with two Lorentzian functions.

*Raman Spectroscopy of TAC and Tape samples*: Raman spectroscopy of the mechanically-exfoliated PtSe$_2$ flakes and TAC films was carried out using a confocal WITec Raman microscope alpha300 R equipped with a 532 nm excitation laser. The Raman measurement was performed in air under ambient conditions with a laser power of about 1 mW, 100x objective lens and a dispersive grating with 1800 l/mm.

*XPS of LPE PtSe$_2$.* Highly concentrated LPE-PtSe$_2$ dispersions were dropped on SiO$_2$ wafer and left to dry at room temperature. The XPS Versa Probe by Physical Electronics was used to acquire all XPS spectra. The system uses the monochromatic Al Kα line (1486.7 eV) with a spot size of 100 µm. The samples are mounted electrically isolated to have a floating potential. Electrical charging of the sample surface was neutralised by using a combination of Ar$^+$-ion and electron flux. First, wide-scan surveys of the whole energy region were performed, and subsequently high-resolution scans of the individual orbitals of interest including Platinum, Selenium, Carbon, and Oxygen. The binding energy scale was referenced to the adventitious carbon 1s core level at 284.8 eV.

*XPS of bulk PtSe$_2$:* X-ray photoelectron spectroscopy was performed using an ESCAProbeP spectrometer (Omicron Nanotechnology Ltd, Germany) with a monochromatic aluminium X-ray radiation source (1486.7 eV). Wide-scan surveys of all elements were performed, with subsequent high-resolution scans. The samples were placed on a conductive carrier made from a high-purity silver bar. An electron gun was used to eliminate sample charging during measurement (1–5 V).

## *Acknowledgement*

C.B. acknowledges the German research foundation DFG under Emmy-Noether grant BA4856/2-1 and Jana Zaumseil for access to the infrastructure of the Applied Physical Chemistry institute. G.S.D. acknowledges support from the European Union's Horizon 2020 research and innovation programme under grant agreement No 785219 (Graphene Flagship). Z.S. acknowledges the Czech Science Foundation (GACR No. 20-16124J). N.M. acknowledges support from Science Foundation Ireland (15/SIRG/3329, 12/RC/2278_P2)

## *Additional information*

Supplementary information is available.

*Competing financial interests*

The authors declare no competing financial interests.

*Figures*:

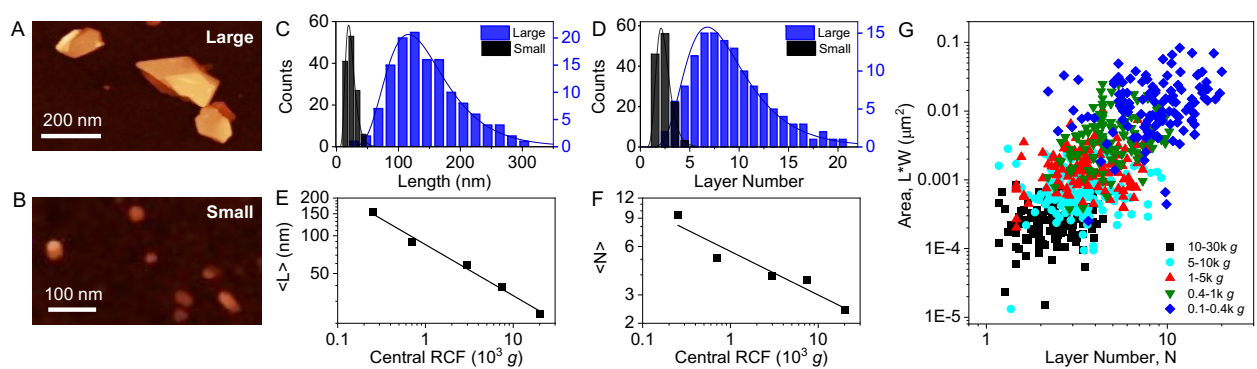

**Figure 1. Microscopic characterisation of liquid phase-exfoliated and size-selected PtSe$_2$ nanosheets**. A, B) Representative AFM image of PtSe$_2$ nanosheets isolated at 0.1-0.4 k *g* (A) and 10-30 k *g* (B), C-D) Histograms for lateral size distribution <L> (C) and layer number (D) of nanosheets isolated at 0.1-0.4 k *g* (blue) and 10-30 k *g* (black), E-F) Mean length <L> (E) and layer number <N> (F) of nanosheets isolated in individual fractions as a function of central centrifugal force g, G) Overall PtSe$_2$ nanosheet population in the dispersions shown as nanosheet area plotted vs. layer number. Each data point represents one nanosheet and colours reflect the fractions isolated in the size selection process.

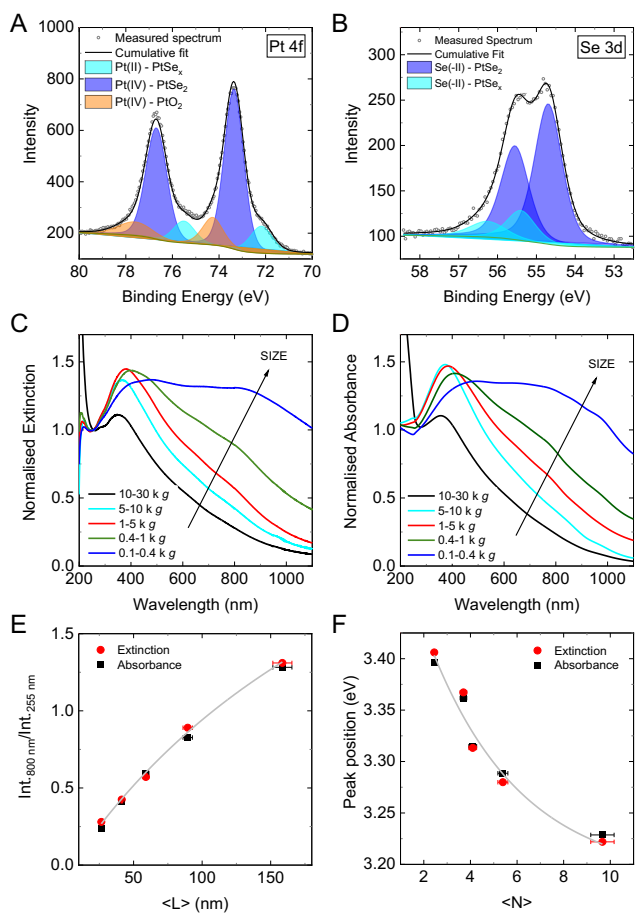

**Figure 2. Spectroscopic characterisation of liquid phase-exfoliated and size-selected PtSe$_2$ nanosheets**. A-B) Fitted XPS platinum 4f (A) and selenium 3d (B) core-level spectra for nanosheets isolated at 10-30 k $g$. C-D) Optical extinction (C) and absorbance (D) spectra normalised to the local minimum, showing size-dependent changes. E) Intensity ratio Ext$_{800}$/Ext$_{255}$ from extinction (red) and intensity ratio Abs$_{800}$/Abs$_{255}$ from absorbance (black) plotted as a function of nanosheet mean length <L> measured by AFM. F) Peak position of the extinction (red) and absorbance (black) spectra (obtained from second derivatives, see SI) plotted vs. mean nanosheet layer number <N> measured by AFM.

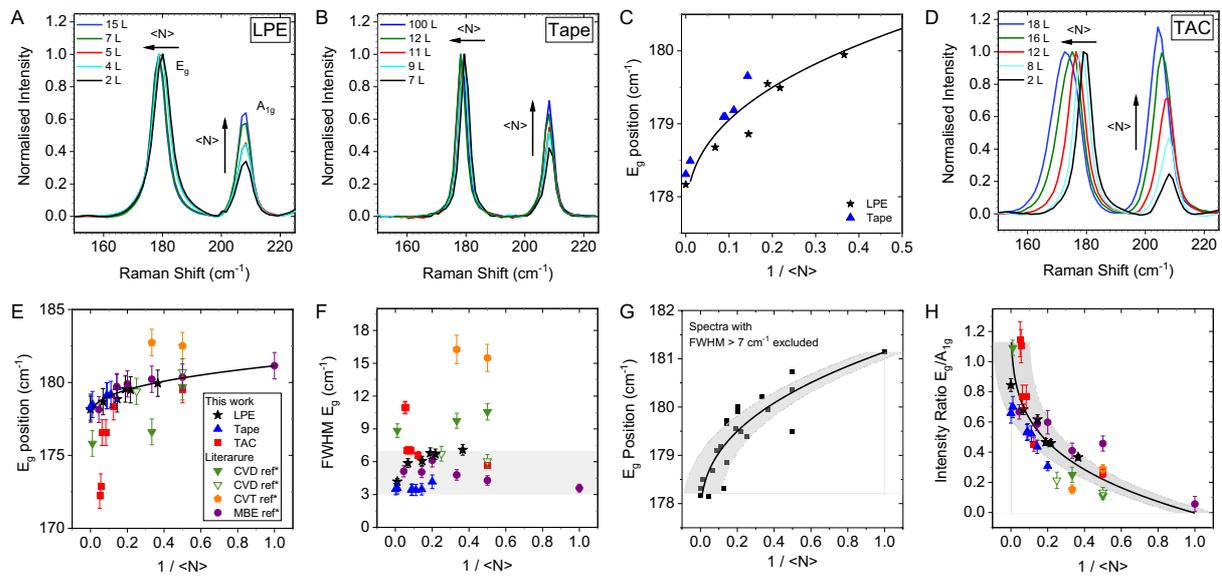

**Figure 3. Raman analysis of various PtSe$_2$ samples produced by top-down and bottom-up techniques.** A-B) Raman spectra of PtSe$_2$ nanosheets of different thickness prepared by top-down exfoliation *via* A) Liquid Phase Exfoliation (LPE), B) Micromechanical Exfoliation (Tape). Thickness-dependent shifts of the E$_g$ mode and changes in the E$_g$/A$_{1g}$ intensity ratio are observed. C) Plot of the E$_g$ positions (extracted from fitting the spectra) as a function of the inverse layer number for LPE and Tape PtSe$_2$ showing a similar scaling. D) Raman spectra of bottom-up grown PtSe$_2$ films of different thickness produced by thermally assisted conversion (TAC). The size-dependent changes are qualitatively similar, albeit they appear more pronounced. E) Plot of E$_g$ position as a function of the inverse layer number for a range of PtSe$_2$ samples including LPE, Tape, TAC and various grown films, where Raman spectra were digitised from literature and fit the same way. A much larger spread in peak positions is observed than in C. F) Plot of full width at half maximum (FWHM) of the E$_g$ mode for the samples in E). G) Same data as in E), but with all samples that exhibited a FWHM > 7 cm$^{-1}$ excluded. A well-defined empirical power law is observed (black line) with most data points falling within a 20% deviation from the fit line (grey box). H) Plot of the intensity ratio E$_g$/A$_{1g}$ as function of inverse layer number (legend as in E). An empirical power law is observed (black fit line) with most data falling within a 20% deviation from the fit line (grey box).